\documentclass[preprint2]{emulateapj}

\shorttitle{High-Resolution X-Ray Spectroscopy of Puppis~A} 
\shortauthors{Katsuda et al.}
\slugcomment{Published in {\em The Astrophysical Journal}, Volume 756, Issue 1, article id. 49 (2012)}

\begin{document}

\title{High-Resolution X-Ray Spectroscopy of the Galactic Supernova 
Remnant Puppis~A with the {\it XMM-Newton} RGS}

\author{Satoru Katsuda\altaffilmark{1}, Hiroshi Tsunemi\altaffilmark{2},
Koji Mori\altaffilmark{3}, Hiroyuki Uchida\altaffilmark{4},
Robert Petre\altaffilmark{5},\\
Shin'ya Yamada\altaffilmark{1}, Hiroki Akamatsu\altaffilmark{6},
Saori Konami\altaffilmark{1, 7}, and Toru Tamagawa\altaffilmark{1}
}
\altaffiltext{1}{RIKEN (The Institute of Physical and Chemical
  Research), 2-1 Hirosawa, Wako, Saitama 351-0198}

\altaffiltext{2}{Department of Earth and Space Science, Graduate School
of Science, Osaka University, 1-1 Machikaneyama, Toyonaka, Osaka,
60-0043, Japan}

\altaffiltext{3}{Department of Applied Physics and Electronic 
Engineering, Faculty of Engineering, University of Miyazaki, 
1-1 Gakuen Kibanadai-Nishi, Miyazaki, 889-2192, Japan}

\altaffiltext{4}{Department of Physics, Kyoto University, 
Kitashirakawa-oiwake-cho, Sakyo, Kyoto 606-8502, Japan}

\altaffiltext{5}{NASA Goddard Space Flight Center, Code 662, Greenbelt
MD 20771}


\altaffiltext{6}{Department of Physics, Tokyo Metropolitan University, 
1-1 Minami-Osawa, Hachioji, Tokyo 192-0397}

\altaffiltext{7}{Department of Physics, Tokyo University of Science, 
1-3 Kagurazaka, Shinjuku-ku, Tokyo 162-8601}

\begin{abstract}
We present high-resolution X-ray spectra of cloud-shock interaction 
regions in the eastern and northern rims of the Galactic supernova 
remnant Puppis~A, using the Reflection Grating Spectrometer onboard 
the {\it XMM-Newton} satellite.  A number of emission lines including 
K$\alpha$ triplets of He-like N, O, and Ne are clearly resolved for the 
first time.  Intensity ratios of forbidden to resonance lines in the 
triplets are found to be higher than predictions by thermal emission 
models having plausible plasma parameters.  The anomalous line ratios 
cannot be reproduced by effects of resonance scattering, recombination, 
or inner-shell ionization processes, but could be explained by 
charge-exchange emission that should arise at interfaces between the 
cold/warm clouds and the hot plasma.  Our observations thus provide 
observational support for charge-exchange X-ray emission in supernova 
remnants.
\end{abstract}

\section{Introduction}

It has long been thought that the X-ray emission from supernova remnants 
(SNRs) is partly due to charge-exchange (CX) processes between neutrals 
and highly-ionized ions \citep[e.g.,][]{Serlemitsos1973}.  
\citet{Wise1989} first performed a detailed theoretical calculation of 
CX-induced X-ray emission from SNRs.  They found that the contribution of 
CX emission is typically 10$^{-3}$ to 10$^{-5}$ compared with that 
of thermal emission (i.e., electron collisional excitation lines).  
This means that CX emission is in general very minor.  However, as the 
authors claimed, CX emission could play an enhanced role where 
neutrals are mixed with shocked hot gas because of hydrodynamic instabilities 
in SNRs and/or inhomogeneities of the interstellar medium.  
Subsequently, \citet{Lallement2004} examined projected emission profiles for 
both CX and thermal emission in SNRs, and noted that CX X-ray emission could 
be comparable with thermal emission in thin layers at the SNR edge.  
It is also pointed out that the relative importance of CX X-ray emission 
to thermal emission is proportional to a quantity, 
$n_\mathrm c$~$V_\mathrm r$~$n_\mathrm e^{-2}$, with $n_\mathrm c$ being 
the cloud density, $V_\mathrm r$ the relative velocity between 
neutrals and ions, and $n_\mathrm e$ the electron density of the 
hot plasma.  Thus, the higher the density contrast 
($n_\mathrm c$/$n_\mathrm e$), the stronger the presence of CX X-ray 
emission would become.

Observational evidence of CX emission from SNRs was found at optical 
wavelengths more than 30\,yrs ago \citep[e.g.,][]{Kirshner1978}, but is 
still lacking in the X-ray domain.  So far, only marginal detections of 
CX X-ray emission were reported in the SMC SNR 1E0102.2--7219 
\citep{Rasmussen2001}, and the Galactic SNR, the Cygnus Loop 
\citep{Katsuda2011}.  The line intensity ratios of 
($n>2 \rightarrow n=1$)/($n=2 \rightarrow n=1$) in H-like O 
(1E0102.2--7219) or He-like O (Cygnus Loop) seem to be higher than 
expectations from thermal emission models, and were interpreted as 
signatures of CX.  However, the CX interpretation is still a matter of 
debate, because {\it Chandra} grating observations of 1E0102.2--7219 
\citep{Flanagan2004} did not confirm the {\it XMM-Newton} result, and 
uncertainties of Fe L line emissivities might cause the spectral anomaly 
in the Cygnus Loop \citep{Katsuda2011}.

One SNR that might show CX X-ray emission is Puppis~A, a large 
($\sim$50$^{\prime}$ in diameter) middle-aged 
\citep[$\sim$3700\,yr:][]{Winkler1988} SNR in our Galaxy.  Its surface 
brightness is the highest among all Galactic SNRs in soft X-rays 
($E<$1\,keV).  The X-ray image is inhomogeneous with particularly bright 
eastern and northern knots (hereafter, BEK and NK).  These regions 
undoubtedly represent strong interactions between the SNR blast wave and 
dense clouds \citep[e.g.,][]{Petre1982,Dubner1988,Hwang2005}, and thus 
they exhibit ideal conditions in which CX X-ray emission could be 
significant \citep{Lallement2004}.  While indications of CX signatures 
have not yet been reported in this SNR, such signatures can be best 
obtained through high-resolution X-ray spectroscopy.

Here, we present high-resolution X-ray spectra of the BEK/NK features 
using {\it XMM-Newton}'s Reflection Grating Spectrometer 
\citep[RGS:][]{denHerder2001}.  From the RGS spectra, forbidden-to-resonance 
line ratios in He$\alpha$ triplets are found to be anomalously enhanced, 
especially at the BEK.  We show that this anomaly could 
\begin{center}
\begin{figure*}
\includegraphics[angle=0,scale=0.5]{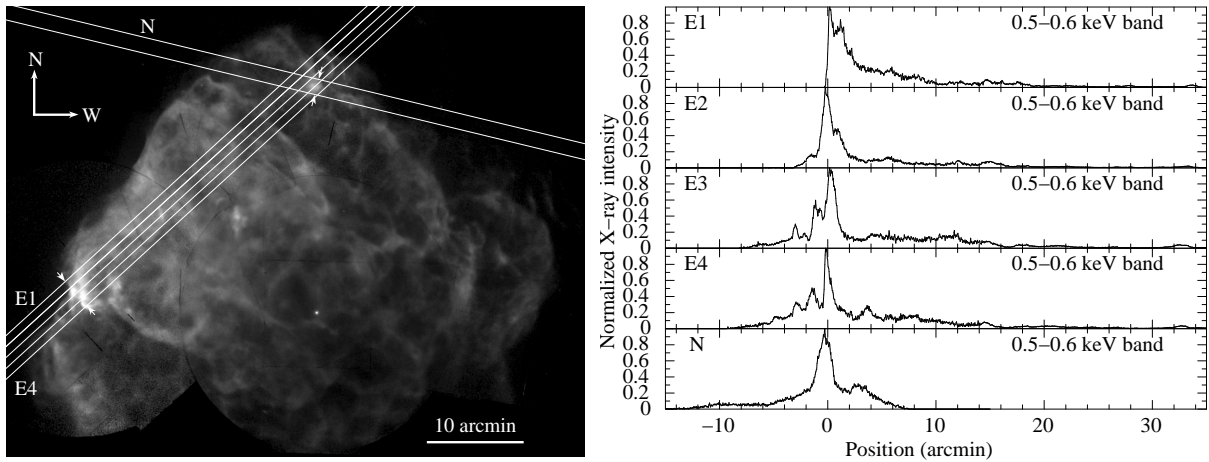}\hspace{1cm}
\caption{Left: Merged {\it XMM-Newton} and {\it Chandra} image in the energy 
  band 0.5--5\,keV.  The intensity scales as the square root of the surface 
  brightness.  The RGS spectral extraction regions are overlaid in white with 
  the region name.  The arrows indicate zero points of projection profiles 
  shown in Fig.~\ref{fig:image} right.
  Right: X-ray emission profiles along the RGS dispersion axis for the five 
  spectral extraction regions.  The zero points (x=0) correspond to the 
  direction from which emission is detected at the nominal (no red/blue shift) 
  wavelength positions on the RGS detectors.
} 
\label{fig:image}
\end{figure*}
\end{center}
be naturally caused by the presence of CX emission.

\section{Observations and Spectral Analysis}

The NK and BEK were observed by {\it XMM-Newton} on 2003 April 17 
(Obs.ID 0150150101) and 2003 May 21 (Obs.IDs 0150150201 and 0150150301), 
respectively, in order to obtain high-resolution spectra using the RGS.  
The dispersion directions of the RGS are 76$^{\circ}$.4 
(NK) and 132$^{\circ}$.3 (BEK) counterclockwise from the north, as shown
in Fig.~\ref{fig:image}, which is an X-ray image of Puppis~A generated 
from existing {\it XMM-Newton} and {\it Chandra} data.  In this paper, 
we mainly focus on the RGS data, while we also utilize data taken by the 
European Photon Imaging Camera \citep[EPIC:][]{Turner2001,Struder2001} 
to support our RGS analyses.  The exposure times after removing intervals 
affected by soft protons are 13.5\,ks and 20.8\,ks for the NK 
and BEK, respectively. All the raw data are processed using version 
11.0.0 of the XMM Science Analysis Software and the latest calibration 
data files available at the time of the analysis. 

Since the RGS is a slitless spectrometer, off-axis sources along the 
dispersion direction are detected at wavelength positions shifted with 
respect to the on-axis source.  Spatial displacement of 
1$^{\prime}$ corresponds to a wavelength shift of 0.138\,\AA (or 
4\,eV/arcmin at 0.6\,keV) for the first spectral order.  Because the BEK 
and the NK are fairly compact (2$^\prime$--3$^\prime$) features with much 
higher surface brightness than their surroundings, the RGS is 
capable of producing high-resolution spectra for them, with an 
order-of-magnitude better resolution than nondispersive CCDs.  
In fact, there are a number of successful RGS observations not only of 
moderately extended SNRs in the LMC/SMC \citep[e.g.,][]{Rasmussen2001}, but 
also of a bright knot along the northwestern rim of SN~1006 \citep{Vink2003}.

As shown in Fig.~\ref{fig:image}, we divide the eastern RGS field of view 
into four sectors spaced by 0$^{\prime}$.8 along the cross-dispersion axis, 
while we extract one spectrum from the 1$^{\prime}$.6-width region for the 
north.  In this way, we mitigate the photon number difference between 
the BEK and the NK.  Also, slicing the BEK region allows us to generate 
more accurate RGS response files than the response file for the entire 
BEK region as we describe below.  
We smooth RGS responses originally designed for point 
sources, based on emission profiles along the dispersion direction with 
the {\tt rgsrmfsmooth} software.  For the input to the software, we arrange 
the X-ray images such that {\it Chandra} covers regions around the BEK/NK 
while {\it XMM-Newton} covers the remaining regions, and we take account of 
vignetting effects of {\it XMM-Newton}'s X-ray telescope.  We generate 
energy-dependent RGS responses using energy-band images in 
0.35--0.4\,keV, 0.4--0.5\,keV, 0.5--0.6\,keV, 0.6--0.7\,keV, 
0.7--0.85\,keV, 0.85--0.97\,keV, 0.97--1.2\,keV, and 1.2--1.5\,keV.  
Figure~\ref{fig:image} right shows the X-ray emission profiles in 
0.5--0.6\,keV for the five regions.  These profiles, smoothed by the 
point spread function of the telescope, become the RGS responses 
themselves for the corresponding energy band.  We note from this figure 
that the contribution from the NK in BEK spectra is almost negligible due 
to vignetting effects (the NK is slightly seen around x$=$33$^{\prime}$ in the
upper four panels in Fig.~\ref{fig:image} right).

For RGS background (BG), we use a blank sky observation (Lockman hole: 
Obs.ID 0147511601) for which the RGS spectrum is free from apparent line 
emission.  In addition, we need to subtract local BG, since the Vela SNR 
is superposed on Puppis~A.  To estimate local BG, we simulate RGS spectra 
based on plasma parameters derived by modeling EPIC spectra outside 
Puppis~A.  The simulated local BG is added to the blank sky spectrum, 
resulting in a total BG.  The total BG is dominated by local BG, and is less 
than 1\% of the source emission.  As for the EPIC BG, we use blank-sky data 
prepared by \citet{Read2003}.  Since X-ray emission from Vela is much 
weaker than the BEK/NK features, especially in the higher energy band, 
we do not take account of local BG for the EPIC spectra.

\begin{figure*}
\begin{center}
\includegraphics[angle=0,scale=0.45]{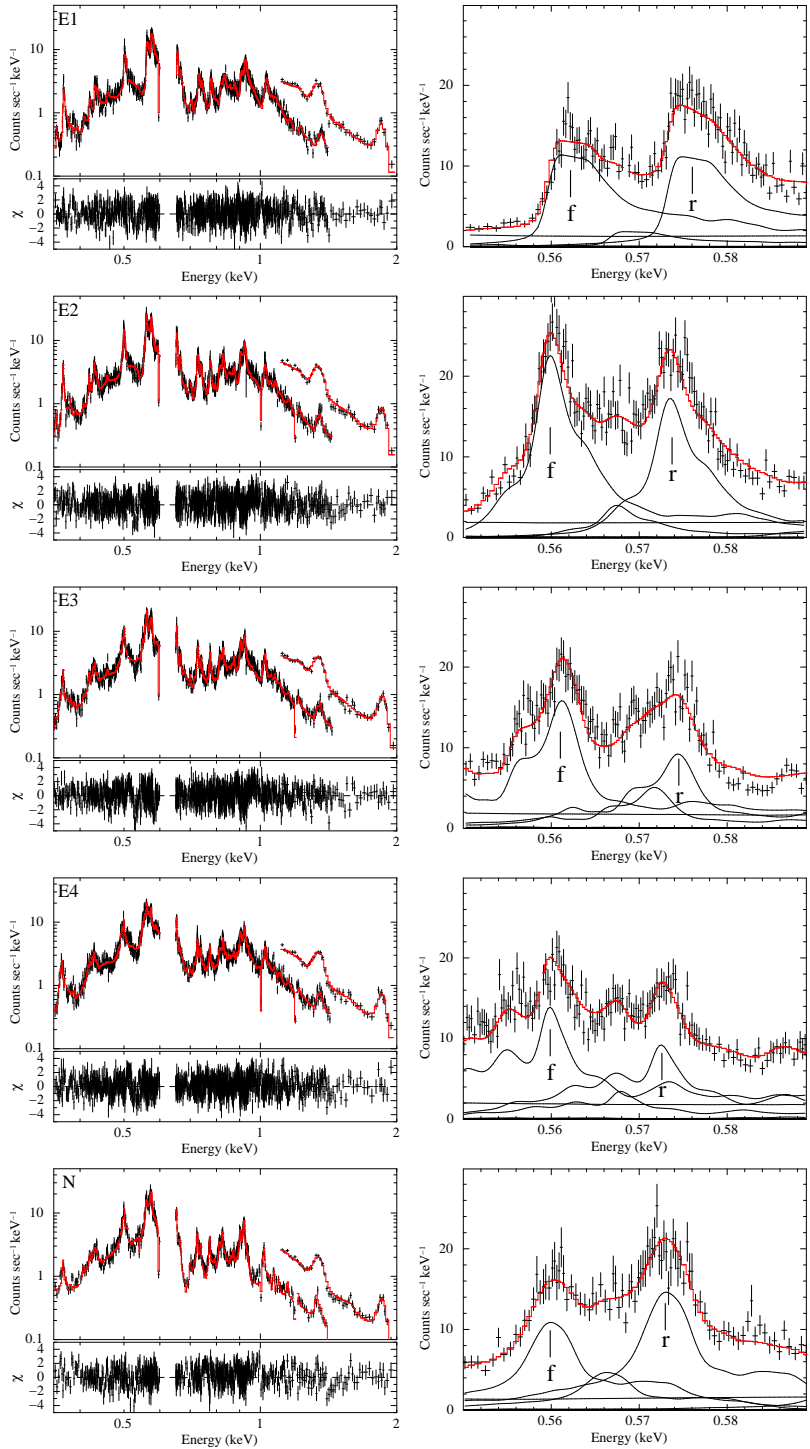}\hspace{1cm}
\caption{Left: {\it XMM-Newton} spectra of the five regions in 
  Fig.~\ref{fig:image}.  The data covering below 0.6\,keV, 0.65--1.5\,keV, 
  and above 1.1\,keV are the first order RGS1+2, the second order RGS1+2, 
  and MOS1+2, respectively.  These data are simultaneously fitted with a 
  phenomenological model (see, text), and the best-fit models are shown 
  in red.  Lower panels show the residuals.  
  Right: Close-up RGS spectra for O He$\alpha$ triplets with individual 
  best-fit model components.  Forbidden and resonance lines are indicated 
  as "f" and "r", respectively.
} 
\label{fig:spec}
\end{center}
\end{figure*}

Figure~\ref{fig:spec} shows BG-subtracted spectra for the five regions.  
To make the best use of {\it XMM-Newton}'s capability, three different 
data sets are analyzed together: the first order RGS spectra below 
0.6\,keV, the second order RGS spectra in the 0.65--1.5\,keV band, and the MOS 
spectra above 1.1\,keV.  Note that the second order RGS spectra, which 
show the best spectral resolution among all the X-ray spectrometers onboard 
{\it XMM-Newton}, do not cover the energy range below 0.65\,keV, and that 
above $\sim$1\,keV the signal-to-noise ratio of the EPIC is much better 
than that of the RGS.  We sum spectra from two co-aligned RGS 
spectrometers, RGS1 and RGS2, to improve photon statistics.  We extract
MOS spectra from the same sliced regions as the RGS, limiting around the 
BEK or the NK.  As expected from previous X-ray observations, the RGS 
spectra are dominated by a number of emission lines, including
C~Ly$\alpha$ @0.375\,keV, N~He$\alpha$ triplet @0.42\,keV,
N~Ly$\alpha$+N~He$\beta$ @0.5\,keV, O~He$\alpha$ triplet @0.57\,keV, 
O~Ly$\alpha$ @0.654\,keV, O~He$\beta$ @0.666\,keV, Fe~L (3s$\rightarrow$2p) 
complex @0.73\,keV, O~Ly$\beta$ @0.775\,keV, Fe~L (3d$\rightarrow$2p) 
complex @0.83\,keV, Ne~He$\alpha$ triplet @0.92\,keV, Mg~He$\alpha$ triplet 
@1.35\,keV, and Si~He$\alpha$ triplet @1.85\,keV.  In particular, as 
shown in the right (and left) panels of Fig.~\ref{fig:spec}, forbidden and 
resonance lines in the He$\alpha$ triplet of O (N and Ne as well) are 
clearly resolved for the first time for the two knots.  We also note that 
this is the first clear detection of C and N lines from Puppis~A.


We fit the spectra with an absorbed bremsstrahlung component plus 49 
Gaussian components.  Free parameters are the H column density $N_\mathrm H$, 
the electron temperature $kT_\mathrm e$, and the normalization of the 
bremsstrahlung component.  We also allow normalizations of most Gaussians to 
vary freely.  Since O~Ly$\gamma$ and O~Ly$\delta$ lines fall in the Fe L 
line forest, we fix the normalization ratios of O~Ly$\gamma$/O~Ly$\beta$ and 
O~Ly$\delta$/O~Ly$\beta$ to 0.3 and 0.1 as are expected at 
$kT_\mathrm{e}\sim$0.5\,keV \citep{Hwang2005}.  We note that the 
$kT_\mathrm{e}$ values of the bremsstrahlung component are mainly determined 
by continuum emission above 1.5\,keV.  Thus, fixing ratios of Gaussian 
normalizations does not affect the $kT_\mathrm{e}$ measurements. Line 
centers of the most prominent 21 Gaussians are left as free parameters, 
while those for C~He$\beta$ and two Fe L lines (G\footnote{We use the same 
labeling for Fe {\scshape XVII} L lines 
as in \citet{Gillaspy2011}.} and D) are systematically shifted with respect 
to those of their neighboring lines (H and E).  Other weak Gaussians' 
centers are fixed to the theoretically expected values \citep{Smith2001}.  
The widths of all Gaussians are fixed to zero, since significant 
broadening is not required from a statistical point of view; note that 
apparent line widths seen in the RGS spectra in Fig.~\ref{fig:spec} are 
due to spatial extent of the source as described in the second paragraph in 
this section.  With this fitting strategy, we obtain fairly good fits for 
all the spectra as shown in Fig.~\ref{fig:spec}.  The fit results are 
summarized in Table~\ref{tab:param}, where we omit weak Gaussians.  

The values of $N_\mathrm H$ are consistent with recent X-ray measurements
\citep{Hwang2005,Hwang2008,Katsuda2010}.  The $kT_\mathrm e$ values are 
somewhat lower than previous results.  However, prominent line intensities, 
which are essential for the discussion below, are not affected by the 
temperature difference, as the fraction of underlying continuum is very 
small.  Line intensity ratios are basically reproduced by thermal emission 
models having plausible plasma parameters \citep[i.e., electron temperature 
of 0.5\,keV and ionization timescale of 
10$^{11}$\,cm$^{-3}$\,s:][]{Hwang2005}.  

On the other hand, as summarized in Table~\ref{tab:lineratios}, the 
intensity ratios of forbidden line to resonance line in He$\alpha$ triplets 
(hereafter, F/R ratios) are inconsistent with model predictions, for which 
we assume a plane-parallel shock model with a distribution of ionization 
timescales appropriate for a plane-parallel shock \citep[{\tt vpshock} 
model in conjunction with augmented NEI version 2.0:][]{Borkowski2001}; 
this emission model is often used to describe X-ray spectra from 
Puppis~A.  In calculating the model F/R ratio, we take a reasonable 
range of plasma conditions in the BEK \citep[$kT_\mathrm{e}$ = 
0.3--0.7\,keV and $n_\mathrm{e} t$ = 
10$^{10}$--5$\times$10$^{11}$\,cm$^{-3}$\,s:][]{Hwang2005}, given that 
the RGS spectra are integrated along the dispersion direction as well as 
the line of sight.  From Table~\ref{tab:lineratios}, we see that the 
F/R-ratio anomaly for O and Ne ions is particularly evident in the E2--E4 
regions compared with that in the E1 and NK regions, while the F/R ratios 
for N ions are marginally consistent among the five regions.  
The discrepancy can 
not be solved by multiple plasma components, since the model predictions 
already consider various (reasonable) plasma conditions.  Although the 
ratios could be reproduced at a very low $n_\mathrm{e}\,t$ value of 
$\sim$10$^{9}$\,cm$^{-3}$\,s, in such a low ionization condition, no 
Ne-like Fe L lines nor H-like Ne lines can be emitted, which is in stark 
contrast to the observed X-ray spectra.  Thus, the F/R ratios cannot be 
reproduced by thermal emission models.

We next investigate whether the anomalous F/R ratio is due to reduction 
of the resonance line or increase of the forbidden line, by comparing the 
observed line intensities with model expectations.  To this end, we fit 
the X-ray spectra (in the E2 region) with an absorbed {\tt vpshock} model, 
excluding the particularly anomalous O He$\alpha$ triplets.  After fitting, 
we recover the O He$\alpha$ triplet lines to compare the data with the 
model.  We find that the intensity of the forbidden line inferred from 
the {\tt vpshock} model is weaker than the data and that the model 
intensity of the resonance line is stronger.  This result suggests both 
reduction of the resonance line and increase of the forbidden line.  
However, we need to be careful about limitations of our spectral modeling.  
This is because the X-ray emission other than the O He$\alpha$ triplets 
may also not be interpreted in the frame of pure thermal emission models, 
due to contamination of CX emission, effects of resonance line scattering, 
or some other processes.  If this is the case, the plasma parameters 
including O abundance, the electron temperature, and the ionization 
timescale inferred by our thermal emission modeling should have some 
systematic uncertainties, leading to an incorrect model intensity of the 
O He$\alpha$ triplet.  Therefore, it is difficult to infer the model 
O He$\alpha$ intensity accurately enough at this point.  We will revisit 
this issue in our future work that will include developments of more 
sophisticated emission models.


\section{Discussions}

We have presented high-resolution X-ray spectra of the BEK/NK features in 
the Galactic SNR Puppis~A.  The forbidden and resonance lines in the 
He$\alpha$ triplets are clearly resolved, and their intensity ratios 
(F/R) are found to be generally higher than predictions from thermal 
plasma models.  This anomaly is particularly evident in the BEK, while the 
RGS spectrum of the NK as well as the {\it Einstein} FPCS spectra of the 
northeastern portion \citep{Winkler1981a,Winkler1981b} are closer to
predictions of thermal models.  

There are two ways to enhance the F/R ratio: (1) an increase of the forbidden 
line flux or (2) a reduction of the resonance line flux.  Possible mechanisms 
for the former are either H-like$\rightarrow$He-like recombination/CX 
processes or inner-shell ionization of Li-like ions due to the higher 
statistical weight.  As for the latter, we consider resonance-scattering 
effects; only resonance lines are scattered out of the line of sight 
because of their higher optical depths.  Among several SNRs for which 
high-resolution spectroscopy has been performed, anomalous F/R ratios 
have only been reported in two LMC SNRs, DEM~L71 \citep{vanderHeyden2003} 
and 0506--68 a.k.a N23 \citep{Broersen2011}.  In these cases, the authors 
attributed the anomaly to recombination and/or resonance scattering, paying 
little or no attention to CX.  Below we discuss which mechanism mainly works 
for the case of Puppis~A.

\begin{deluxetable*}{llccccc}
\tabletypesize{\tiny}
\tablecaption{Spectral-fit parameters}
\tablewidth{0pt}
\tablehead{
\colhead{Component}&\colhead{Parameter}&\multicolumn{5}{c}{Region}}
\startdata
 & & E1 & E2 & E3 & E4 & N\\
\hline
Absorption & $N_{\mathrm H}$ ($10^{21}$cm$^{-2}$) & 2.70$\pm$0.01 & 2.80$\pm$0.01 & 2.80$\pm$0.01 & 2.85$\pm$0.01 & 2.58$\pm$0.01 \\
\hline 
Bremsstrahlung & $kT$$_{\mathrm e}$ (keV) & 0.32$\pm0.01$ & 0.31$\pm0.01$ & 0.33$\pm0.01$ & 0.32$\pm0.01$ & 0.30$\pm0.01$ \\
& Normalization$^a$ & 5264$\pm$93 & 9124$\pm$133 & 6744$\pm$117 & 7808$\pm$117 & 5397$\pm$90 \\
\hline 
Gaussian:~~C Ly$\alpha$+Si~L+S~L & Center (eV) & 366.8$^{+0.2}_{-0.1}$ & 367.2$^{+0.2}_{-0.1}$ & 366.7$^{+0.4}_{-0.3}$ & 365.4$^{+0.4}_{-0.3}$ & 367.4$^{+0.3}_{-0.8}$ \\
& Normalization$^a$ & 822$\pm$80 & 1486$\pm$118 & 1257$\pm$139 & 1723$\pm$139 & 380$\pm$52 \\
~~~~~~~~~~~~~~\,N He$\alpha$ (f) & Center (eV) & 419.2$^{+0.2}_{-0.5}$ & 419.2$\pm$0.3 & 418.4$\pm$0.5 & 419.7$\pm$0.5 & 420.5$^{+0.8}_{-0.6}$ \\
& Normalization$^a$ & 100$\pm$18 & 201$\pm$26 & 145$\pm$26 & 243$\pm$33 & 82$\pm$16 \\
~~~~~~~~~~~~~~\,N He$\alpha$ (i) & Center (eV) & 426$^b$ & 426$^b$ & 426$^b$ & 426$^b$ & 426$^b$ \\
& Normalization$^a$ & $<$15 & 57$\pm$23 & 72$\pm$24 & $<$57 & 18$\pm$15 \\
~~~~~~~~~~~~~~\,N He$\alpha$ (r) & Center (eV) & 429.7$^{+0.3}_{-0.2}$ & 430.1$^{+0.3}_{-0.2}$ & 429.9$\pm$0.5 & 429.7$\pm$0.5 & 430.6$\pm$0.5 \\
& Normalization$^a$ & 146$\pm$17 & 223$\pm$24 & 169$\pm$23 & 199$\pm$27 & 72$\pm$15 \\
~~~~~~~~~~~~~~\,C Ly$\beta$ & Center (eV) & 434.9$^{+0.7}_{-0.2}$ & 435.1$^{+0.6}_{-0.3}$ & 435.2$^{+0.7}_{-0.6}$ & 434.7$^{+0.7}_{-0.6}$ & 435.2$^{+0.7}_{-0.6}$ \\
& Normalization$^a$ & 60$\pm$14 & 87$\pm$18 & 55$\pm$18 & 87$\pm$22 & 51$\pm$13 \\
~~~~~~~~~~~~~~\,C Ly$\gamma$ & Center (eV) & 459$^b$ & 459$^b$  & 459$^b$  & 459$^b$  & 459$^b$ \\
& Normalization$^a$ & 25$\pm$8 & 31$\pm$10 & 24$\pm$10 & 14$\pm$12 & $<8$ \\
~~~~~~~~~~~~~~\,N Ly$\alpha$+He$\beta$ & Center (eV) & 499.4$\pm$0.2 & 500.2$^{+0}_{-0.2}$ & 499.8$^{+0.2}_{-0.3}$ & 499.6$^{+0.2}_{-0.3}$ & 499.6$^{+0.5}_{-0.4}$ \\
& Normalization$^a$ & 143$\pm$9 & 254$\pm$13 & 222$\pm$12 & 252$\pm$14 & 162$\pm$11 \\
~~~~~~~~~~~~~~\,O He$\alpha$ (f) & Center (eV) & 559.8$^{+0.4}_{-0.1}$ & 560.7$\pm$0.1 & 560.5$\pm$0.5 & 560.2$\pm$0.5 & 561.2$\pm$0.2 \\
& Normalization$^a$ & 705$\pm$27 & 1167$\pm$36 & 1042$\pm$36 & 1119$\pm$38 & 417$\pm$23 \\
~~~~~~~~~~~~~~\,O He$\alpha$ (i) & Center (eV) & 566.8$^{+1.0}_{-0.9}$ & 568.3$^{+0.3}_{-1.1}$ & 570.7$^{+0.4}_{-0.7}$ & 568.5$^{+0.4}_{-0.7}$ & 567.5$^{+0.4}_{-1.1}$ \\
& Normalization$^a$ & 107$\pm$24 & 190$\pm$30 & 298$\pm$31 & 255$\pm$34 & 168$\pm$22 \\
~~~~~~~~~~~~~~\,O He$\alpha$ (r) & Center (eV) & 573.1$\pm$0.2 & 574.1$^{+0.3}_{-0.2}$ & 573.4$\pm$0.1 & 572.9$\pm$0.1 & 574.3$^{+0.2}_{-0.3}$ \\
& Normalization$^a$ & 601$\pm$25 & 771$\pm$30 & 527$\pm$31 & 646$\pm$33 & 492$\pm$21 \\
~~~~~~~~~~~~~~\,O Ly$\alpha$ & Center (eV) & 652.7$^{+0.5}_{-0.1}$ & 653.4$\pm$0.2 & 653.0$\pm$0.2 & 652.8$\pm$0.2 & 653.4$\pm$0.4 \\
& Normalization$^a$ & 979$\pm$38 & 1223$\pm$47 & 1154$\pm$46 & 1329$\pm$51 & 631$\pm$36 \\
~~~~~~~~~~~~~~\,O He$\beta$ & Center (eV) & 664.4$\pm$0.8 & 665.0$^{+0.3}_{-0.7}$ & 665.4$^{+1.3}_{-1.4}$ & 666.0$^{+1.3}_{-1.4}$ & 667.0$\pm$1 \\
& Normalization$^a$ & 161$\pm$24 & 285$\pm$29 & 197$\pm$27 & 136$\pm$29 & 110$\pm$17 \\
~~~~~~~~~~~~~~\,O He$\gamma$ & Center (eV) & 699.0$^{+0.9}_{-0.8}$ & 700.7$\pm$0.6 & 698.9$^{+1.1}_{-0.8}$ & 698.2$^{+1.1}_{-0.8}$ & 700.7$\pm$1.1 \\
& Normalization$^a$ & 68$\pm$8 & 74$\pm$10 & 65$\pm$11 & 74$\pm$13 & 42$\pm$9 \\
~~~~~~~~~~~~~~\,Fe L (G+H) & Center$^{c}$ (eV) & 724.9$^{+0.4}_{-0.1}$ & 725.1$^{+0.2}_{-0.4}$ & 724.9$^{+0.1}_{-0.3}$ & 725.2$^{+0.1}_{-0.3}$ & 725.2$^{+0.1}_{-0.4}$ \\
& Normalization$^a$ & 247$^{+16}_{-15}$ & 362$^{+21}_{-20}$ & 331$^{+19}_{-18}$ & 332$^{+20}_{-19}$ & 156$^{+13}_{-12}$ \\
~~~~~~~~~~~~~~\,Fe L (F) & Center (eV) & 739.3$^{+0.8}_{-0.5}$ & 738.5$^{+0.1}_{-0.7}$ & 739.3$^{+0.9}_{-0.3}$ & 738.1$^{+0.9}_{-0.3}$ & 738.9$^{+0.8}_{-0.3}$ \\
& Normalization$^a$ & 89$\pm$11 & 156$\pm$13 & 94$\pm$12 & 99$\pm$12 & 65$\pm$8 \\
~~~~~~~~~~~~~~\,O Ly$\beta$ & Center (eV) & 773.5$^{+0.2}_{-0.6}$ & 773.8$\pm$0.3 & 773.3$^{+0.4}_{-0.6}$ & 772.9$^{+0.4}_{-0.6}$ & 773.7$\pm$0.5 \\
& Normalization$^a$ & 151$\pm$7 & 187$\pm$8 & 160$\pm$7 & 169$\pm$8 & 97$\pm$6 \\
~~~~~~~~~~~~~~\,Fe L (D+E) & Center$^{c}$ (eV) & 810.6$\pm$0.8 & 810.1$\pm$0.2 & 810.0$\pm$0.4 & 810.0$\pm$0.4 & 810.1$\pm$0.7 \\
& Normalization$^a$ & 74$^{+10}_{-9}$ & 110$^{+11}_{-10}$ & 98$^{+11}_{-10}$ & 90$^{+11}_{-10}$ & 50$^{+7}_{-6}$ \\
~~~~~~~~~~~~~~\,Fe L (C) & Center (eV) & 826.6$^{+0.4}_{-1.0}$ & 826.1$\pm$0.5 & 825.6$\pm$0.5 & 826.1$\pm$0.5 & 826.2$^{+0.3}_{-0.6}$ \\
& Normalization$^a$ & 105$\pm$6 & 173$\pm$0.8 & 135$\pm$7 & 152$\pm$8 & 78$\pm$5 \\
~~~~~~~~~~~~~~\,Ne He$\alpha$ (f)+Fe L & Center (eV) & 902.0$\pm$0.6 & 903.3$^{+0.5}_{-0.2}$ & 902.6$^{+0.4}_{-0.2}$ & 902.6$^{+0.4}_{-0.2}$ & 903.5$^{+0.6}_{-0.4}$ \\
& Normalization$^a$ & 122$\pm$6 & 185$\pm$7 & 156$\pm$6 & 175$\pm$7 & 75$\pm$4 \\
~~~~~~~~~~~~~~\,Ne He$\alpha$ (i)+Fe L & Center (eV) & 915$^b$ & 915$^b$ & 915$^b$ & 915$^b$ & 915$^b$ \\
& Neormalization$^a$ & 63$\pm$6 & 53$\pm$7 & 68$\pm$6 & 50$\pm$7 & 18$\pm$4 \\
~~~~~~~~~~~~~~\,Ne He$\alpha$ (r)+Fe L & Center (eV) & 921.2$^{+0.3}_{-0.5}$ & 921.6$^{+0.1}_{-0.3}$ & 921.2$^{+0.3}_{-0.4}$ & 921.8$^{+0.3}_{-0.4}$ & 922.4$\pm$0.2 \\
& Normalization$^a$ & 141$\pm$6 & 224$\pm$7 & 168$\pm$6 & 188$\pm$7 & 111$\pm$5 \\
~~~~~~~~~~~~~~\,Ne Ly$\alpha$+Fe L & Center (eV) & 1023.6$^{+1.2}_{-1.4}$ & 1022.5$^{+0.5}_{-1.2}$ & 1020.5$^{+0.2}_{-1.4}$ & 1022.1$^{+0.2}_{-1.4}$ & 1020.4$^{+1.1}_{-0.8}$ \\
& Normalization$^a$ & 83$\pm$5 & 97$\pm$6 & 96$\pm$5 & 102$\pm$6 & 35$\pm$3 \\
\hline 
\multicolumn{2}{l}{$\chi^{2}$/d.o.f.} & 1423/968 & 1740/1151 & 1698/1114 & 1664/1181 & 949/612 \\
\enddata
\label{tab:param}
\tablecomments{$^a$In units of $10^{-4}$\,photons\,cm$^{-2}$\,s$^{-1}$. \\$^b$Fixed values.\\$^c$Line centers of H and E are shown, while those of G and D are systematically shifted by +2\,eV from H and E, respectively.}
\end{deluxetable*}

\begin{deluxetable*}{lcccccc}
\tabletypesize{\tiny}
\tablecaption{Line intensity ratios}
\tablewidth{0pt}
\tablehead{
\colhead{Element}&\colhead{Thermal predictions$^{a}$}&\multicolumn{5}{c}{Region}}
\startdata
& & E1 & E2 & E3 & E4 & N\\
\hline
N F/R & 0.38--0.52 & 0.68$\pm$0.15 & 0.90$\pm$0.15 & 0.86$\pm$0.19 & 1.22$\pm$0.23 & 1.14$\pm$0.33 \\
O F/R & 0.42--0.56 & 1.17$\pm$0.07 & 1.51$\pm$0.08 & 1.98$\pm$0.13 & 1.73$\pm$0.11 & 0.85$\pm$0.06 \\ 
Ne$^{b}$ F/R & 0.44--0.71 & 0.87$\pm$0.06 & 0.83$\pm$0.04 & 0.93$\pm$0.05 & 0.93$\pm$0.05 & 0.68$\pm$0.05 \\
N F/I  & 5.8--6.4 & $>$7.9 & 3.49$\pm$1.46 & 2.02$\pm$0.76 & $>$3.7 & 4.48$\pm$3.73 \\
O F/I & 4.2--5.1 & 6.58$\pm$1.49 & 6.14$\pm$0.99 & 3.49$\pm$0.38 & 4.39$\pm$0.60 & 2.48$\pm$0.35 \\
Ne$^{b}$ F/I & 3.3--4.9 & 1.92$\pm$0.20 & 3.45$\pm$0.47 & 2.30$\pm$0.22 & 3.46$\pm$0.50 & 4.01$\pm$0.87 \\
(N Ly$\alpha$ + N He$\beta$) / N He$\alpha$ & 0.12--0.9 & 0.58$\pm$0.07 & 0.53$\pm$0.05 & 0.58$\pm$0.06 & 0.54$\pm$0.06 & 0.94$\pm$0.13 \\
O Ly$\alpha$ / O He$\alpha$ & 0.01--1.03 & 0.69$\pm$0.03 & 0.57$\pm$0.03 & 0.62$\pm$0.03 & 0.66$\pm$0.03 & 0.59$\pm$0.04 \\
Ne$^{b}$ Ly$\alpha$ / Ne$^{b}$ He$\alpha$ & 0--0.83 & 0.25$\pm$0.02 & 0.21$\pm$0.01 & 0.24$\pm$0.01 & 0.25$\pm$0.02 & 0.17$\pm$0.02 \\
Fe L (F+G+H)/C & 1.4--2.6 & 3.19$\pm$0.26 & 2.98$\pm$0.14 & 3.15$\pm$0.23 & 2.84$\pm$0.21 & 2.83$\pm$0.26 
\enddata
\label{tab:lineratios}
\tablecomments{$^a$Predictions by the {\tt vpshock} model at 
($kT_\mathrm{e}$, $n_\mathrm{e} t$) $=$ 
\citep[0.3--0.7\,keV, 10$^{10}$--5$\times$10$^{11}$\,cm$^{-3}$\,s][]{Hwang2005}.  
$^b$The observed values could be affected by Fe L lines.}
\end{deluxetable*}
\vspace{2cm}

Effects of resonance scattering can be evaluated for each emission line
from its optical depth, $\tau$ \citep[e.g.,][]{Kaastra1995}.  We calculate
the value of $\tau$ using the BEK's typical plasma parameters; i.e., $\sim$0.5 
solar abundances, $kT_\mathrm e$$\sim$0.5\,keV, 
$n_\mathrm{e}t$$\sim$10$^{11}$\,cm$^{-3}$\,s, and the hydrogen column 
density of the X-ray--emitting plasma $\sim2.4\times$10$^{19}$\,cm$^{-2}$ 
from $n_\mathrm e$$\sim$4\,cm$^{-3}$ 
\citep{Arendt2010} multiplied by an assumed plasma depth of 3.2\,pc, which 
corresponds to the 3$^{\prime}$ size of the BEK at a distance of 2.2\,kpc 
\citep[][we refer to this distance hereafter]{Reynoso2003}.  
To derive oscillator strengths and ion fractions, we utilize the data bases 
from CHIANTI \citep{Dere2009} and SPEX \citep[specifically, the {\tt neij} 
code:][]{Kaastra1996}, respectively.  Then, assuming no microturbulence, 
we calculate $\tau$ and the escape probability, $p$.  The escape 
probabilities for resonance lines in He$\alpha$ triplets of N, O, and Ne 
are obtained to be 1, 0.92, and 0.92, respectively, while those for 
forbidden lines are all unity because of their very small oscillator 
strengths.  Therefore, we find that effects of resonance line 
scattering likely play a modest role, if any, in explaining the observed 
F/R ratios for N and O in all the five regions, while the F/R anomaly for 
Ne could be explained by the effects.  
Whereas optical depths could differ with changing the assumed 
plasma depth and microturbulence velocity, the difference seems small in 
our case.  For example, in order to reproduce 
the observed F/R ratio in the E3 region, we need a plasma depth of 77\,pc
(or 2$^{\circ}$) which is obviously too long for this region.  Also, a 
microturbulence velocity of 100\,km\,s$^{-1}$ (which is statistically 
allowed by our RGS spectral analysis) would reduce $\tau$ to $\sim$0.12 
and increase $p$ to $\sim$0.95 for a reasonable plasma depth of 3.2\,pc.
It should be noted that the 
Fe L line ratio, (F+G+H)/C, should be more sensitive to resonance 
scattering than the O {\scshape VII} F/R ratio: we expect 
$\tau$$\sim$0.02 and $p$$\sim$0.99 for F+G+H and $\tau$$\sim$0.43 and 
$p$$\sim$0.84 for C for the plasma conditions in the BEK.
We find in Table~\ref{tab:lineratios} that the observed (F+G+H)/C 
ratios are indeed slightly higher than those of the {\tt vpshock} 
prediction.  While the measured line ratios suggest the presence of 
resonance scattering for the Fe L lines, the degree of the effect is not 
as significant as what is expected from the anomaly of O {\scshape VII} 
F/R ratios.  Therefore, we are required to examine other possibilities.


X-ray--emitting recombining plasmas have been recently found in several 
SNRs \citep[e.g.,][and references therein]{Yamaguchi2009,Uchida2012}.  
The recombining plasma is characterized by strong radiative recombination 
continua (RRC) and enhanced Ly$\alpha$/He$\alpha$ ratios.  We find that 
signatures of RRCs (i.e., strong recombination edges) are not evident in the 
RGS and MOS spectra in Fig.~\ref{fig:spec} (and also {\it Suzaku} XIS 
spectra which are not shown in this paper), although the Fe L and/or other 
line emission might make it difficult for us to detect such signatures.  
Also, the Ly$\alpha$/He$\alpha$ ratios for N, O, and Ne in the five 
regions are all well within thermal predictions as can be seen in 
Table~\ref{tab:lineratios}, while the BEK regions show somewhat larger 
ratios than those in the NK.  In addition, recombination processes would 
enhance the Fe L line ratio (F+G+H)/(C+D+E) to $>$25 \citep{Liedahl1990}, 
which is inconsistent with our RGS measurements of 1.2--2.4 in the BEK and 
the NK.  Furthermore, our spectral fitting of the combined RGS and MOS 
spectra with a recombining plasma model \citep[the {\tt cie} model in 
SPEX:][]{Kaastra1996} failed to reproduce the entire X-ray spectrum; the 
model requires too low an electron temperature to explain emission above 
1\,keV.  These investigations indicate that the plasma here is not 
recombining.

Signatures of inner-shell ionization processes of Li-like ions can
be found as Li-like satellite lines as well as an enhanced 
forbidden-to-intercombination (F/I) ratio compared with collisionally 
excited emission \citep[e.g.,][]{Porquet2010}.  In the RGS spectra, 
there is no indication of satellite lines, however.  Also, the measured 
F/I ratios of He$\alpha$ triplets in Table~\ref{tab:lineratios} are 
all marginally consistent with thermal expectations; we need to take 
account of considerable systematic uncertainties on the intensity of 
the intercombination line due to its weakness compared to the 
surrounding forbidden and resonance lines.  
These facts led us to conclude that inner-shell ionization 
processes are not working efficiently in this region.  We note that, 
whereas line emission from Li-like O is seen in the far-ultraviolet 
spectrum of the BEK \citep{Blair1995}, the abundance of Li-like O would 
be small in the X-ray--emitting region.

We next assess the feasibility of the CX scenario by calculating the 
expected CX flux, following \citet{Lallement2004}.  The volume emissivity 
of CX is expressed as 
$P_\mathrm {CX}$ = $\sigma_\mathrm {CX}$~$n_\mathrm H$~$n_\mathrm i$~$V_\mathrm r$.
We let $\sigma_\mathrm {CX}$ be the CX cross section between neutral H 
and ions of interest, $n_\mathrm H$ the neutral H density, 
$n_\mathrm i$ the ion density, and $V_\mathrm r$ the relative H-ion 
velocity.  We first consider the O {\scshape VII} forbidden line, so 
that $\sigma_\mathrm {CX}$ is 3.3$\times$10$^{-15}$\,cm$^{-2}$ 
\citep{Bodewits2007}.  The value of $n_\mathrm H$ can be taken from the 
density of cold/warm clouds immersed in X-ray--emitting plasma, which 
is $\sim$50\,cm$^{-3}$ \citep{Teske1987}.
The value of $n_\mathrm i$ is the density of O {\scshape VIII} in the hot 
plasma, which is $\sim$2$\times$10$^{-4}$\,cm$^{-3}$ for the BEK 
\citep{Hwang2008,Arendt2010}.  We assume that 
$V_\mathrm r$ = 500\,km\,s$^{-1}$, roughly the shock velocity that can 
produce the X-ray--emitting plasma in the BEK.  These parameters give 
$P_\mathrm {CX}$$\sim$1.7$\times$10$^{-9}$\,photons\,cm$^{-3}$\,s$^{-1}$.  
The CX-emitting volume is calculated by the thickness of the CX layer, 
$l_\mathrm {CX}$, times the interface area between the clouds and the hot 
plasma.  The value of $l_\mathrm{CX}$ is equated to the mean free path for 
H-proton CX.  Thus, $l_\mathrm{CX}$ is of the order of 
1/$\sigma$$n_\mathrm p$, with $\sigma$ being the H-proton CX cross section 
\citep[10$^{-15}$\,cm$^{-2}$:][]{McClure1966} and $n_\mathrm p$ being 
the proton density in the plasma \citep[4\,cm$^{-3}$:][]{Arendt2010}.  
Considering that only $\sim$30\% neutral H can charge-exchange before 
getting collisionally ionized \citep{Lallement2004} for the BEK's plasma 
condition, we obtain an effective CX thickness of
$\sim$7.5$\times$10$^{13}$\,cm.  An assumed cylindrical interface with a 
diameter of $3^{\prime}$ would give a surface area of the CX layer of 
$\sim$3$\times$10$^{37}$\,cm$^{2}$ for each region, resulting in a 
CX-emitting volume of $\sim$2.3$\times$10$^{51}$\,cm$^{3}$.  
With $P_\mathrm{CX}$ and the emitting volume estimated, and an assumption 
of a uniform $P_\mathrm{CX}$, we obtain an unabsorbed CX flux for the 
O {\scshape VII} forbidden line of 
$\sim$6$\times$10$^{-3}$\,photons\,cm$^{-2}$\,s$^{-1}$ 
for each region in the BEK.  Next, we repeat the calculation for 
forbidden lines of He-like N and Ne.  Using the relation that 
$\sigma_\mathrm{CX}$ is proportional to the atomic number, Z, and 
the ion number densities of $n_\mathrm{N VI}\sim8\times10^{-6}$\,cm$^{-3}$ 
and $n_\mathrm{Ne IX}\sim4\times10^{-5}$\,cm$^{-3}$ \citep[which are based 
on plasma parameters in][]{Hwang2005}, we estimate the unabsorbed CX 
fluxes of the N {\scshape VI} and Ne {\scshape IX} forbidden lines to 
be $\sim3\times$10$^{-4}$\,photons\,cm$^{-2}$\,s$^{-1}$ and 
$\sim2\times$10$^{-3}$\,photons\,cm$^{-2}$\,s$^{-1}$, respectively.

We should keep in mind that the uncertainty of the CX flux estimate is 
fairly large.  For example, we implicitly assume that hydrogen in the 
cold/warm clouds is entirely neutral due to rapid recombination caused by 
radiative cooling in the dense cloud.  The hydrogen must be partially 
ionized, however, resulting in a lower CX flux.  On the other 
hand, hot neutral hydrogen formed after H-proton CX reactions, 
dust destruction, and unshocked cold neutrals going through the 
(collisionless) shock would all provide additional neutrals in the hot 
plasma, increasing the CX flux.  Also, the area of the interface region 
is quite uncertain.  These factors suggest that our CX flux estimate 
should only be considered accurate to an order of magnitude.

To compare the expected CX flux with our measurements in 
Table~\ref{tab:param}, we need to correct for the spatial-integration 
factor.  This is because the expected flux is estimated for the BEK 
feature, whereas the fluxes in Table~\ref{tab:param} are integrated along 
the RGS dispersion direction.  To estimate the correction (reduction) 
factors, we extract EPIC spectra from 3$^{\prime}$-long regions including 
the BEK, and normalize the RGS flux to equalize the EPIC flux in each 
region.  In this way, the correction factors are derived to be 3.7--4.5 for 
the four BEK regions.  The fluxes of the N, O, and Ne He$\alpha$ 
forbidden lines are, respectively, derived to be 
$\sim$5$\times$10$^{-3}$\,photons\,cm$^{-2}$\,s$^{-1}$, 
$\sim$3$\times$10$^{-2}$\,photons\,cm$^{-2}$\,s$^{-1}$, and 
$\sim$5$\times$10$^{-3}$\,photons\,cm$^{-2}$\,s$^{-1}$ for 
each region in the BEK.  We thus find order-of-magnitude agreement with the 
expected CX fluxes.  Given the considerable uncertainty of the expected 
CX flux, we cannot rule out the CX scenario with this level of agreement.
In addition, there are two pieces of observational evidence supporting the 
CX scenario.  First, optically-emitting clouds, which we think are the major 
electron donors, are seen only in the BEK, whereas such clouds are not 
detected in the other regions including the NK and the {\it Einstein} FPCS 
field of view.  Second, the N/O and Ne/O abundance ratios based on CX fluxes 
in the He$\alpha$ triplets for the BEK region are consistent with previous 
X-ray measurements (N/O$\sim$1 and Ne/O$\sim$2, respectively).  Here, we 
applied the same method used in \citet{Liu2011} who modeled He$\alpha$ 
triplets with a thermal emission model plus Gaussians to represent the CX 
emission for the starburst galaxy M82.  In this context, we conclude that 
CX emission is a promising mechanism for explaining the anomalous F/R 
ratios observed in Puppis~A.

One might expect to find other spectral signatures of CX in addition to the 
He$\alpha$ line ratios.  However, CX spectral properties are strongly 
dependent on $V_\mathrm r$ and the target neutrals 
\citep[e.g.,][]{Beiersdorfer2001,Beiersdorfer2003}, and investigation of 
CX emission is still ongoing.  Therefore, further discussion calls 
for more sophisticated CX emission modeling, which is beyond the scope of 
this paper and is left as future work.

\section{Conclusions}

High-resolution X-ray spectra of the cloud-shock interaction regions, the 
BEK and the NK, in Puppis~A have revealed anomalous He$\alpha$ triplet 
ratios: in particular, the O He$\alpha$ F/R line ratios are found to be 
$\sim$2 in the BEK.  This anomalous ratio can be naturally interpreted 
as the result of the presence of CX emission, although resonance-scattering 
effects might be non-negligible.  

Future observations with the non-dispersive Soft X-ray Spectrometer 
\citep[SXS:][]{Mitsuda2010} onboard the {\it Astro-H} satellite 
\citep{Takahashi2010} will allow for further tests for the CX scenario 
and other possibilities.  While the RGS can perform high-resolution 
spectroscopy only for bright knotty features, the SXS with expected 
spectral resolution of 4\,eV FWHM and spatial resolution of 1$^{\prime}$ 
HPD \citep{Serlemitsos2010} will enable us to reveal a spatial distribution
of the F/R ratios of the knotty features and their surroundings.  The CX 
scenario can then be examined by seeing if there is a spatial correlation 
between the F/R ratios and optical clouds which we here consider to be 
main electron donors.  We expect that other large Galactic SNRs, especially 
the Cygnus Loop introduced in Section 1, would be also promising sites to 
check the presence of CX X-ray emission.


\acknowledgments

We would like to thank Prof.\ H. Tanuma for fruitful discussions about 
the laboratory experiments of CX X-ray emission.  We also thank the 
referee for numerous comments which improved the quality of the paper.
S.K.\ and S.Y.\ are supported by the Special Postdoctoral Researchers 
Program in RIKEN.  This work is partly supported by a Grant-in-Aid for 
Scientific Research by the Ministry of Education, Culture, Sports, 
Science and Technology (23000004).



\end{document}